\documentclass[12pt]{article}
\usepackage[dvips]{epsfig}
\begin{document}
\begin{flushright}
OHSTPY-HEP-T-99-031
hep-th/9904027
\end{flushright}
\vspace{20mm}
\begin{center}
{\LARGE The Perils of `Soft' SUSY Breaking }
\\
\vspace{20mm}
{\bf F.Antonuccio, O. Lunin,  and  S.Pinsky \\}
\vspace{4mm}
Department of Physics,\\ The Ohio State University,\\ Columbus, OH 43210, USA\\
\vspace{4mm}
\end{center}
\vspace{10mm}
\begin{abstract}
We consider a two dimensional SU($N$) gauge theory coupled to an adjoint
Majorana fermion, which
is known to be supersymmetric for a particular value of fermion mass.  We
investigate the
`soft' supersymmetry breaking of the discrete light cone quantization (DLCQ)
of this theory. 
There are several DLCQ
formulations of this theory currently in the literature and they naively
appear to behave
differently under `soft' supersymmetry breaking at finite resolution. We
show that all these formulations nevertheless
yield identical bound state masses 
in the decompactification limit of the light-like circle.
Moreover, we are able to show that the supersymmetry-inspired
version of DLCQ (so called `SDLCQ') provides
the best rate of convergence of DLCQ bound state
masses towards the actual continuum values, 
except possibly near or at the critical fermion mass. 
In this last case, we discuss improved extrapolation 
schemes that must supplement the DLCQ algorithm in order
to obtain correct continuum bound state masses.
Interestingly, when we truncate the Fock space
to two particles, the SDLCQ prescription presented here
provides a scheme for improving the rate of convergence of the
massive t'Hooft model. Thus the supersymmetry-inspired SDLCQ
prescription is applicable to theories {\em without} supersymmetry. 

\end{abstract}
\newpage

\baselineskip .25in

\section{Introduction}
Over the last several years we have learned a great deal about
supersymmetric gauge theories following the discovery
of dualities between string/M-theory  and
supersymmetric gauge theories
\cite{bfs97,sus97,mald97}. Recently
this has been extended to conformal field theories without
supersymmetry \cite{klt99}.
Evidently, it would be desirable to have a deeper understanding
of supersymmetry breaking in order to bridge
the gap between the formulation of physics in a supersymmetric world, and
its more realistic counterpart, where no such symmetry is manifest.
One straightforward approach is to start with
a supersymmetric formulation, and then proceed to break
supersymmetry `softly' by adding appropriate mass terms.

The context within which we will consider this is a theory that has been
well studied
before: two dimensional SU($N$) gauge theory coupled to an adjoint Majorana
fermion
\cite{bdk93}. Interestingly, this theory is known to exhibit
supersymmetry at a particular value of the fermion mass, $m=m_{SUSY}$
\cite{kut93}.
This is believed to be a theory with two parameters $g$ and $m$, 
both of which
have the
dimensions of mass. Since the only $g$ dependence is an overall $g^2$
factor in the
Hamiltonian the theory depends on one dimensionless parameter $X={m^2 \pi
\over g^2 N}$
and therefore all the bound state masses, in units of ${g^2 N \over \pi}$,
must be
determined in terms of the one parameter $X$. In this work, we provide
evidence that,
while this viewpoint is still correct, 
there is still scope for an additional operator
(and associated coupling constant) that may be introduced
to improve convergence of
the DLCQ bound state masses towards their actual continuum values.
Of course, these continuum masses will be unaffected
by the presence of such an operator, but a judicious choice of
coupling will serve to improve the rate of convergence
of our numerical results.

Naively, when one adds a `soft' breaking term to the two 
DLCQ formulations of the theory, 
we appear to arrive at different spectra. 
The two formulations we are alluding to are the
`Principle Value' (PV) and `Supersymmetric Discrete light Cone 
Quantization' (SDLCQ), and
are discussed
in detail below. The question is whether these two spectra are truly
different or
simply rescalings of the same spectrum which become identical in the
continuum limit. 
It would be very good news if they were in fact the same because
the PV prescription is
generally accepted as correct (\cite{pab85,bpp98}), while the SDLCQ approach
is known to converge more rapidly in general. 

Actually, to understand the relation between these two schemes,
it is helpful to present a formulation that {\em interpolates} between the 
PV and SDLCQ prescriptions by introducing an
additional operator \cite{alp99} and associated 
coupling constant that we will call $Y$.
In particular, $Y=0$ will correspond to the PV prescription,
while $Y=1$ will imply the SDLCQ prescription. Intermediate
values for $Y$ will correspond to a `mixture' of the two schemes.  
By diagonalizing the DLCQ Hamiltonian matrix, and extrapolating to 
the continuum limit, we are able to solve for bound state 
masses and wave functions at
different values of the fermion mass parameter $X$ and coupling 
constant $Y$. 

We shall show that the continuum bound state masses are independent
of the coupling $Y$, as expected from a scheme independent
prescription, 
although the {\em rate of convergence}
towards the actual continuum mass will be significantly affected
by our choice for $Y$. In fact, it will turn out
that the value for $Y$ that arises naturally in the regularization
of supersymmetric theories (i.e. $Y=1$) provides the 
best convergence towards actual continuum masses.
Thus, the supersymmetric formulation of DLCQ (SDLCQ),
corresponding to $Y=1$ -- first
highlighted in the work \cite{mss95} -- yields 
a method for improving numerical convergence of DLCQ bound state masses
even for theories {\em without supersymmetry}.   
 
To show this, we study the DLCQ bound state integral equations
at high resolution, which is made possible by
truncating the Fock space to two particles. 
In particular, we show that the SDLCQ approach converges more uniformly
and rapidly for all values of $X$ that are
sufficiently far from the critical value $X=0$. 

We remark that 
at the supersymmetric point $X=1$, the SDLCQ prescription
preserves supersymmetry even in the discretized theory.  
The advantages of such an approach have
been  exploited in a study of a wide class of supersymmetric gauge theories
in two
\cite{alp98a,alp98b,alpp98,alpp99,alp99} and three dimensions
\cite{alp99b}. 

\section{\bf Formulations of The Theories}

In this section we will consider the  formulations of $1+1$ dimensional QCD
coupled to
adjoint Majorana fermions having arbitrary mass (see for example
\cite{bdk93}) in the light
cone gauge $A^+=0$.  After eliminating
non-physical degrees of freedom by solving constraint equations,
the light--cone
components of total momentum are found to be:
\begin{eqnarray}
\label{momenta}
P^+&=&\int dx^- Tr(i\psi\partial_-\psi),\\
P^-&=&\int dx^-Tr\left(-\frac{im^2}{2}\psi\frac{1}{\partial_-}\psi-
\frac{g^2}{2}J^+\frac{1}{\partial_-^2}J^+\right) \nonumber \\
&=&\frac{m^2}{2}\int_0^\infty\frac{dk}{k}b^\dagger_{ij}(k)b_{ij}(k)+
\frac{g^2N}{\pi}\int_0^\infty\frac{dk}{k}\int_0^k dp \frac{k}{(p-k)^2}
b^\dagger_{ij}(k)b_{ij}(k) +\nonumber\\
&&\frac{g^2}{2\pi}\int_0^\infty
dk_1dk_2dk_3dk_4\left(\delta(k_1+k_2-k_3-k_4)
A(k)b^\dagger_{kj}(k_3)b^\dagger_{ji}(k_4)b_{kl}(k_1)b_{li}(k_2)+\right.\\
&&\left.\delta(k_1+k_2+k_3-k_4)B(k)(b^\dagger_{kj}(k_4)b_{kl}(k_1)b_{li}(k_2)b_{
ij}(k_4)-
b^\dagger_{kj}(k_1)b^\dagger_{jl}(k_2)b^\dagger_{li}(k_3)b_{ki}(k_4))
\right)\nonumber
\end{eqnarray}
with
\begin{eqnarray}
A(k)=\frac{1}{(k_4-k_2)^2}-\frac{1}{(k_1+k_2)^2},\nonumber\\
B(k)=\frac{1}{(k_3+k_2)^2}-\frac{1}{(k_1+k_2)^2}.
\end{eqnarray}
Here $x^\pm =(x^+ \pm x^-)/\sqrt2 $ and
$J^+_{ij}=2\psi_{ik}\psi_{kj}$ is the longitudinal component of  the
fermion current. To avoid
introducing an additional mass scale in the theory we will write this in
term of mass operators:
$M^2=2P^+P^-$. It is well known that at the special value of fermionic mass
(namely
$m_{SUSY}^2=g^2N/\pi$) this system is supersymmetric \cite{kut93}. We will
use a dimensionless
mass  parameter $X=\frac{\pi m^2}{g^2N}$, and the supersymmetric point is
$X=1$ and the masses
of all bound states will be quoted in units of
$g^2 N/\pi$.The supercharge is given by
\begin{eqnarray}
\label{sucharge}
Q^-&=&2^{1/4}\int dx^-tr(2\psi\psi\frac{1}{\partial_-}\psi). \nonumber \\
&=&\frac{i2^{-1/4}g}{\sqrt{\pi}}\int_0^\infty dk_1dk_2dk_3
\delta(k_1+k_2-k_3)
\left(\frac{1}{k_1}+\frac{1}{k_2}-\frac{1}{k_3}\right)\times\nonumber\\
&\times&\left(b^\dagger_{ik}(k_1)b^\dagger_{kj}(k_2)b_{ij}(k_3)+
b^\dagger_{ij}(k_3)b_{ik}(k_1)b_{kj}(k_2)\right),
\end{eqnarray}
\noindent Using the anticommutator at equal $x^+$:
\begin{equation}
\{\psi_{ij}(x^-),\psi_{kl}(y^-)\}=\frac{1}{2}\delta(x^- -y^-)
\end{equation}
it can be checked that at $m=m_{SUSY}$ the SUSY algebra
$\{Q^-,Q^-\}=2\sqrt{2}P^-$ is satisfied. In the DLCQ approximation the
system lives in a $x^-$ box
of length $L$ and one has to sums over discrete variables
$k^+ \ne 0$ instead of integrations in the above formulas.  For periodic
boundary conditions (BC),
$k^+= n\pi/L $ where $n=1,2,\dots, K$ and
$K$ is called the resolution.

One formulation of DLCQ which we will denote as the
principal value (PV) prescription \cite{tho74}, treats the singularities of the
Hamiltonian using a PV prescription and can be formulated using
either anti- periodic or periodic BC. The anti-periodic boundary
condition must break the
supersymmetry at finite resolution because the fermions and bosons are in
different Fock
sectors. The PV prescription with periodic BC could in principle give
supersymmetric results
at finite resolution, although this is not the case.
In the PV prescription the
supersymmetry at $X=1$
is restored only in the decompactification limit ($K\rightarrow\infty$). This
restoration was
shown in \cite{bdk93} \footnote{They find that convergence is slower for
period BC}. The
Hamiltonian for this formulation  will be referred to as
$P^-_{PV}$.

The prescription  that preserves supersymmetry at finite
resolution will be called
SDLCQ. In SDLCQ one simply uses DLCQ to calculate the supercharges and then
uses
the super charges to calculate the Hamiltonian and longitudinal momentum
operator \cite{mss95}. Here we must use periodic BC because the supercharge
$Q^-$ is cubic in the fields, while the supercharge $Q^+$ is
quadratic.

The SUSY algebra is reproduced at a special value of
fermion mass and at
every finite
resolution the supercharge matrices give a representation of the super
algebra.
Both SDLCQ and $P^-_{PV}$ at $X=1$ give the same results as the resolution goes
to infinity \cite{alp98b} \footnote{However SDLCQ converges much faster}.
We now want to add
identical `soft' SUSY breaking terms (mass terms) to
these theories and study the resulting non-supersymmetric theory. Since
we already have a
mass term in $P^-_{PV}$ this only requires  varying $X$, but for
SDLCQ this means explicitly adding a mass term.

It is very instructive to actually do the numerical calculation
differently and introduce a third formulation, $P_{SUSY}$ which includes
both SDLCQ
and $P^-_{PV}$ .  We
have found the operator which is the difference between the
SDLCQ and the PV
formulation \cite{alp99}
\footnote{To date we have only found this operator for this particularly
simple theory but it
should be possible to find it for other theories as well.
 The calculation of this operator involves a careful study of the
intermediate zero modes that
contribute to the square of the supercharge}. Thus if we add this operator to
the PV Hamiltonian it is now supersymmetric at every resolution and
produces exactly the same
mass and wave functions as SDLCQ.  In the large $N$ approximation the
operator take the form.

\begin{equation}
\label{operator}
\frac{g^2NK}{\pi}\sum_n \frac{1}{n^2}
B^\dagger_{ij}(n)B_{ij}(n).
\end{equation}

Numerically, this operator does not alter the actual 
continuum values observed in the PV approach
when $X=1$. In our
numerical formulation of
$P^-_{SUSY}$ we included this operator with an adjustable coupling $Y$. We
can now think of
$P^-_{SUSY}$ as a single theory in the coupling constant space
$(X,Y)$. The formulation we called PV corresponds 
to setting $Y=0$ and allowing $X$ to be arbitrary,
while the prescription we call SDLCQ corresponds to
setting $Y=1$. In the following, we will present results for the lightest
bosonic bound states
as a function of $X$ and $Y$. For a few values of $X$ and $Y$ we will 
truncate the Fock space 
to allow only two
particles Fock states, which will permit
us to investigate the t'Hooft equation for higher resolutions than
would otherwise be possible.

\section{\bf `Soft' SUSY Breaking}

Our investigation of this theory indicates that at $X=1$ ( the
supersymmetric value of the fermion mass) the lightest
fermionic and bosonic bound states are degenerate with
continuum masses approximately $M^2= 26$ \cite{bdk93,alp98b}.  
Using $P^-_{SUSY}$ we arrive at the same conclusion
for any value of $Y$.

Boorstein and Kutasov \cite{kub94} have investigated `soft' supersymmetry
breaking for small values of
this difference,
$X-1$ and they found that the degeneracy between the fermion and boson
bound state masses is broken
according to
\begin{equation} M^2_F(X) - M^2_B(X)= (1-X) M_B(1)+O((X-1)^3).
\label{linear}
\end{equation}

They calculated these masses using the PV prescription ($Y=0$) with
anti-periodic BC and found
very good agreement with the theoretical prediction.  We have compared this
theoretical prediction
at $Y=1$ and we  find
 that eq (\ref{linear}) is  very well satisfied. At resolution $K=5$, for
example, the slope is 4.76
and the predicted slope
$M_B(1)$ is 4.76. The indication is that this result is true for any value
of $Y$.
\begin{figure}[h]
\begin{center}
\epsfig{file=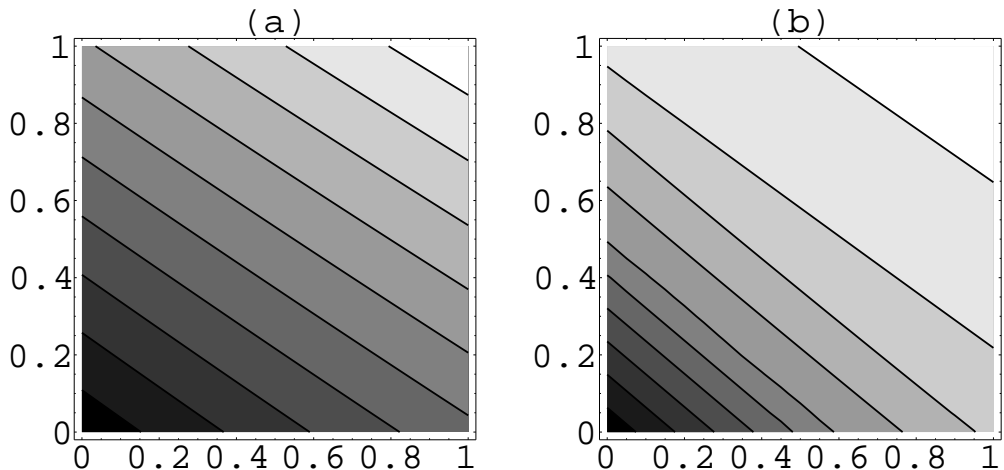, width=14cm}
\end{center}
\caption{(a) The contour plots of
$Y=Y(X)$ for the mass squared of the lowest bound state in units of $g^2 N/
\pi$ as a function of
$X=m \pi /g^2 N$ and Y (b)The contour plots of
$Y=Y(X)$ for the mass squared of the second lowest bound state in units of
$g^2 N/
\pi$ as a function of
$X=m \pi /g^2 N$ and Y (b) }
\end{figure}

In Fig. 1 we show the contour plots of the mass squared $M^2$ 
of the two lightest bosonic bound
states as a function of $X$
and $Y$ at resolution $K=10$. These contours
are lines of constant mass squared. Selecting a particular value of the
mass of the first bound
state then fixes a particular contour in Fig. 1a as a contour
of fixed mass, which we can write as
$Y=Y_p(X)$. 

Interestingly, constructing the same contour plot
for the next to lightest bosonic bound state -- see
Fig. 1b -- yields contours that have approximately
the same functional dependence implied by Fig. 1a.
In fact, one obtains approximately the same contour
plots for the next twenty bound states (which is as far as we checked).
The simple conclusion is that the coupling $Y$ which
represents the strength of the additional operator 
affects all bound state masses more or less equally.
This in turn suggests that at finite resolution, we can
smoothly interpolate between different values of fermion
mass $X$, and different prescriptions specified by the
coupling $Y$, without affecting too much the actual numerical spectrum.
Of course, in the decompactification limit $K \rightarrow \infty$,
such a dependence on $Y$ disappears, due to scheme independence.

Since the lightest bosonic bound state is primarily a
two particle state it is reasonable to truncate the 
Fock basis to  two
particle states. This will permit 
very high resolutions, which will be needed
to carefully scrutinize any possible discrepancies between 
the two versions of
'soft' symmetry breaking presented here. 
In fact, we are able to study the theory
for $K$ up to 800. The
mass of the lowest state as a function of the resolution for various
values of $X$ and $Y$ are
shown in Fig. 2. Each converging pair 
of lines -- which extrapolate the actual data points -- in Fig. 2 
corresponds to different
values of fermion mass $X$. The top
upper curve in each pair runs through data points
that were calculated via SDLCQ (i.e. $Y=1$), while the 
lower corresponds to the PV (i.e. $Y=0$) prescription
commonly adopted in the literature. We
find that each pair of
curves converge to the same point at infinite resolution, 
although this may not be completely obvious
for the lowest pair in the
figure (corresponding to the critical mass $X=0$). 

Away from
$X=0$, the SDLCQ formulation is fitted with a linear function of $1/K$, 
while the PV formulation is fit
with a polynomial of
$1/K^{2\beta}$, where $\beta$ is the solution of $1-X/2=\pi \beta 
Cot(\pi
\beta)$ \cite{van96}. It
now appears that SDLCQ not only provides more rapid convergence 
for supersymmetric models, but
also for the
massive t'Hooft model, which is not supersymmetric.  
For the massless case, the situation is reversed;
the SDLCQ formulation
converges slower. It is fit by a polynomial in $1/Log(K)$ and gives the
same mass at infinite
resolution as the PV formulation. This behavior 
may be understood from the observation that
the wave function of this
state does not vanish at
$x=0$. We have looked closely at `small' masses,
such as $X=.1$, and one finds that both PV and SDLCQ
vary as a polynomial
in $1/K^{2\beta}$ at large resolution. Thus careful extrapolation
schemes must be adopted at small masses.

We therefore conclude that the continuum of 
regularization schemes that interpolate smoothly between 
the SDLCQ and PV prescriptions -- which we characterized
by the parameter $Y$ -- yield the same continuum
bound state masses, although the rate of
convergence of the DLCQ spectrum may be altered significantly. 
This implies that
the contour plots observed in Fig. 1 eventually
approach lines parallel to the $Y$ axis, and the 
sole dependence on the parameter $X$ is recovered.

\begin{figure}[h]
\begin{center}
\epsfig{file=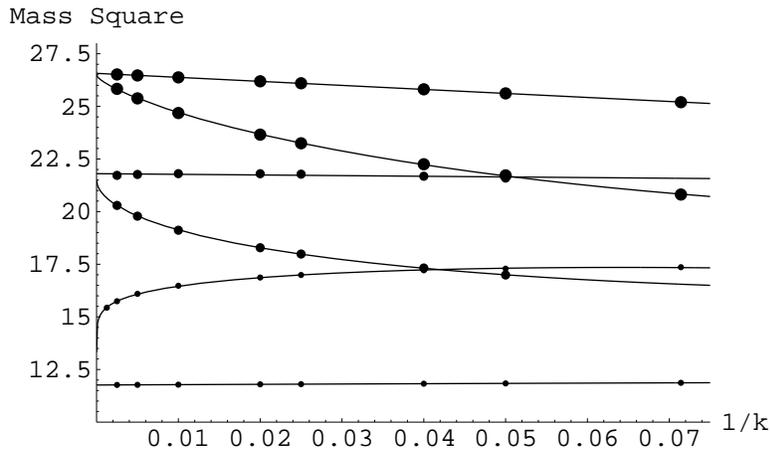}
\end{center}
\caption{Mass of the of the lowest bound state in units of $g^2 N/ \pi$
calculated in the t'Hooft
model. The top pair is at $X=1$, the second is at $X=.5$, 
and the bottom pair is at $X=0$ }
\end{figure}

\section{\bf Discussion}

The two dimensional gauge theory of adjoint Majorana fermions has been
studied extensively
\cite{bdk93,kut93,alp98b,kub94,anp98} and is known to be a theory with an
overall mass scale $g^2$,
and one real coupling -- the mass of the fermion -- which we
write as $X$ in our notation.
When one adds a `soft' supersymmetry breaking term, the supersymmetric
(SDLCQ) and principle
value (PV) prescriptions for regulating
the Coulomb singularity appear to give different bound state
masses at finite resolution.

We observed at finite resolution that these different 
bound state masses may be smoothly connected -- in an approximate sense --
by introducing a new operator, and an associated coupling $Y$,
and then varying the couplings $X$ and $Y$ along an 
appropriately chosen contour.

By truncating the Fock space to two particles,
we were able to study the DLCQ bound state equations up to
$K=800$, which we summarized in Fig. 2.
We concluded that after carefully extrapolating
the data, the different prescriptions yielded
identical continuum bound state masses.
Moreover, we observed that the SDLCQ prescription
improved convergence for sufficiently large values
of fermion mass.

Interestingly, since the 
two-body equation studied here 
for the adjoint fermion model
is simply the t'Hooft equation with
a rescaling of coupling constant, 
we have arrived at an
alternative prescription
for regulating the Coulomb singularity
in the massive t'Hooft model
that improves the rate of convergence towards
the actual continuum mass.
Thus, a prescription that
arises naturally in the study of supersymmetric
theories is also applicable in the study of
a theory without supersymmetry.
We believe that this idea deserves to be exploited
further in a wider context of theories.
In particular, it is an open
question whether this procedure could provide a sensible approach to 
regularizing softly broken gauge theories with bosonic degrees
of freedom, and in higher dimensions.

In any case, it appears that the special cancellations 
afforded by supersymmetry -- especially in
the context of DLCQ bound state calculations --
might have scope beyond the domain
of supersymmetric field theory. This would be a crucial first step
towards a serious non-perturbative study of
theories with broken supersymmetry.

{\bf Acknowledgments}

\noindent The work was supported in part by a grant from
the US Department of Energy. The authors are grateful
for useful discussion with Brett Van de Sande.

\end{document}